\documentstyle[11pt,aasms4]{article}


\begin{document}

\title{COBE's Galactic Bar and Disk}
\author{H.T. Freudenreich}
\affil{Hughes STX}
\authoraddr{Code 685.9, NASA/Goddard Space Flight Center, Greenbelt, MD 20771\\
freudenreich@gsfc.nasa.gov}

\begin{abstract}

A model of the bar and old stellar disk of the Galaxy has been derived from 
the survey of the Diffuse Infrared Background Experiment (DIRBE) of the Cosmic 
Background Explorer at wavelengths of 1.25, 2.2, 3.5, and 4.9 $\mu$m. It
agrees very well with the data, except in directions in which the 
near-infrared optical depth is high. Among the conclusions drawn from the
model: The Sun is located approximately 16.5 pc above the midpoint of the
Galactic plane. The disk has an outer edge four kpc from the Sun, and is 
warped like the \ion{H}{1} layer. It has a central hole roughly the diameter 
of the inner edge of the ``three-kiloparsec" molecular cloud ring, and within 
that hole lies a bright, strong, ``early-type" bar, tilted approximately 
$14^\circ$ from the Sun-Galactic center line. The model has 47 free parameters.
The model is discussed in detail and contour plots and images of the residuals
presented.

\end{abstract}


\clearpage

\section{Introduction}

Only in recent years has it become accepted that our Galaxy is barred.
For decades the prevailing assumption had been that the central concentration
of stars, hidden by intervening dust, could be described by a spheroid in 
which the density of stars fell as a power of the distance from the center. So 
much was suggested by the distribution of globular clusters in the Galaxy, and
by the structure of elliptical galaxies and the apparent shape of the central 
bulges of many spiral galaxies viewed edge-on. What little observational 
evidence there was, obtained by counting stars in low-extinction windows 
toward the inner Galaxy, did not disprove this notion. Some examples of 
spheroidal bulge models may be found in de Vaucoleurs \& 
Pence\markcite{deVbulge} (1978), Bahcall \& Soneira \markcite{BS} (1980), 
Kent\markcite{kentspher} (1992), and Zhao\markcite{zhao} (1996).

Meanwhile, radioastronomy had been accumulating evidence that the inner part of
the Galaxy is less neatly arranged. The radial velocities of gas in the inner
few kiloparsecs was found to be inconsistent with travel in circular orbits. 
In some directions, the velocity is so great that the gas was proposed to lie 
in an ``expanding three-kiloparsec arm" (Oort, Kerr \& 
Westerhout\markcite{oort} 1958) between us and the center; and in general, its
motion seemed predominantly outward, as if the gas were being driven by 
titanic explosions near the Galactic center (Burbridge \& Hoyle 
\markcite{burb}1963), but this notion lost favor as the improbably vast 
energy, and driven mass, required of such explosions came to be better 
understood (Sanders \& Prendergast\markcite{san} 1974).

An alternative explanation is that the gas is moving in non-circular orbits
because the potential it lies in is asymmetric, perhaps due to a bar, as
de Vaucoleurs\markcite{deVbar} (1964) suggested. Other came to a similar 
conclusion, among them Liszt \& Burton\markcite{liszt} (1980) and Gerhard 
\& Vietri\markcite{ger} (1986). Binney et al.\markcite{bin} (1991) 
took this analysis further. From the kinematics of \ion{H}{1}, CO and CS, they 
deduced the presence of a bar whose near end lay in the first Galactic 
quadrant, tilted $16^\circ\pm2^\circ$ from the line joining Sun and Galactic
center.

By this time information from stars in the central part of the Galaxy had
begun to arrive, mostly from observations in the mid- and near-infrared. 
Harmon \& Gilmore\markcite{har} (1986) constructed a picture of the shape of 
the stellar bulge by selecting Mira variables, whose period is a function 
of luminosity, from the IRAS point source catalog on the basis of color. They 
found the bulge to be flattened in $Z$ and, though they did not remark on
it, the outer contours of their Fig. 3a show the bulge to be broader at 
positive longitudes. Nakada et al.\markcite{na} (1992), Whitelock \& 
Catchpole\markcite{white} (1992), and Weinberg \markcite{wein}(1992) also 
found this effect in the distribution of IRAS Miras, and noted it was 
consistent with the appearance of a bar with near end in the first quadrant. 
More evidence came through the near-infrared. Matsumoto et al.\markcite{mat} 
(1982) had mapped the surface brightness of the bulge at 2.4 $\mu$m; the 
pattern of the longitudinal asymmetry of this map was read by Blitz \& 
Spergel\markcite{blitz} (1991) as the clear signature of a bar.  

The Diffuse Infrared Background Experiment (DIRBE) of the Cosmic Background
Explorer (COBE) mapped the entire sky in 10 wavelength bands, including four 
in the near-infrared (Boggess et al.\markcite{bog} 1990). Weiland et 
al.\markcite{wei} (1994) corrected the DIRBE 2.2 $\mu$m map for extinction 
and subtracted an extrapolated projection of the Galactic disk, leaving behind
a boxy, very bar-like object that seems to loom forward into the first 
quadrant. Dwek et al.\markcite{dwek} (1995) applied a variety of bar and 
spheroid models to this same data, and concluded that the bulge is definitely
a bar, with a tilt angle of $20^\circ\pm10^\circ$. 

Gravitational microlensing has recently been proposed as a means of studying
the inner Galaxy. Results from the OGLE project regarding the bar are still 
preliminary (Paczy{\'n}ski\markcite{pac} et al. 1994), but the stellar 
database assembled for the project has allowed Stanek\markcite{sta} et al. 
(1997) to discover a bar tilted $20^\circ$ to $30^\circ$.

Freudenreich\markcite{paper1} (1996, henceforth Paper I), presented a model
of the Galactic disk derived from the DIRBE observations at 1.25, 2.2, 3.5 and 
4.9 $\mu$m, where the surface brightness is dominated by red giants. The 
DIRBE photometric bands approximate the standard $J$, $K$, $L$ and $M$ bands. 
(Here the ``D" subscript used to denote DIRBE in Paper I will be omitted.) The 
model presumed that mature stars and the diffuse component of the interstellar 
dust each form a homogeneous, possibly warped, disk; if the structure of the 
Galaxy appears different in different near-infrared bands, it is due solely to 
the wavelength-dependent extinction and emission by dust. The model fit the 
data very well, except in the directions of nearby molecular clouds, which a 
smooth dust model cannot account for, and toward the inner $3-4$ kiloparsecs 
of the Galaxy, which had been intentionally masked out to avoid the bulge. In 
this paper I have used the same data but expanded the model to include the 
bulge and the inner part of the disk. The ``young disk" (the ``thin disk" in 
Paper I), the ragged collection of molecular clouds, young stars, and 
associated dust concentrated at the Galactic plane, makes penetration of the 
inner part of the Galaxy difficult, but in the near-infrared the extinction 
is low enough that, with the proper techniques, we can discern structure to 
within roughly half a kiloparsec of the center. 

Subsequent sections discuss the data reduction and optimization scheme,
the model itself, how well it fits the data, its residuals, and what it tells
us about the Galaxy. There are frequent references to Paper I, in particular 
for a more detailed explanation of the data reduction and a review of the 
model parameters pertaining to the disk, but all the information needed to 
understand the model is included here. Comparison to others work is made in 
the course of discussing the parameters of the model.

\section{Procedure}

\subsection{Data Preparation}

The analysis was performed on cylindrical equal-area projections of the DIRBE 
full-sky surface brightness maps, each map consisting of 391,612 pixels.
The zodiacal light was modeled and removed, and point sources (nearby stars
and a few supergiants) filtered out. (See Paper I for a more detailed 
explanation.) At low latitudes, $|b|\!<\!20^\circ$, unresolved sources 
dominate the DIRBE surface brightness, but at higher latitudes the 
contribution of point sources cannot be neglected. To account for the deleted 
point sources, the simplifying assumption was made that all stars within 
a ``deletion distance" $D_\nu$ of the Sun were filtered out, which in effect 
places us at the center of a star-free bubble. There are three deletion 
distances: for the $J$, the $K$, and the $L$ and $M$ bands. The $M$ band 
does not have its own deletion distance because at high Galactic latitudes
it is seriously contaminated by zodiacal light or artifacts of its removal.

Maps of the near-infrared colors were then used to identify and mask 
anomalously red areas of the sky, most of which are strongly associated with 
molecular clouds. The mask derived from the $K-L$ color has been retained 
from Paper I, without, of course, the rectangle covering the Galactic bulge. 
In the inner Galaxy, this mask covers everything within approximately three 
degrees of the Galactic equator. A second mask, chosen on the basis of $J-K$ 
color, has been added for use with the $J$-band map only. See 
Fig.~\ref{sample}. (From this point on, the term ``primary mask" refers to 
both of these masks used in conjunction.) The Magellanic Clouds and the heart 
of the Barnard 78 cloud complex ($l\approx1^\circ,\; b\approx4^\circ$) were 
excluded by rectangular masks.

Shrinking the masked area would certainly introduce more contamination by 
young-disk features, which the robust fitting procedure might or might not be 
able to accomodate, yet there are reasons to consider doing so. One is that 
we could approach the Galactic center more closely. Another is that the mask 
is an implicit parameter of the fit, and by using more than one we can gauge 
our sensitivity to it. As further justification, we may note that a line of 
sight at low latitude toward the inner Galaxy can accumulate a great deal of 
reddening solely from diffuse dust, which the model is intended to account 
for, in an averaged way. Therefore I created a ``minimal" mask. For 
$|l|\!>\!60^\circ$ or $|b|\!>\!8^\circ$, it is identical to the primary mask
for all four bands, but interior to this only the Barnard 78 clouds and a
narrow strip ($-1.1^\circ\!<\!b\!<\!0.9^\circ$) along the Galaxy's brightness 
crest, which has been eroded by point-source removal, are excluded.

As a further precaution against zodiacal light residue, a zone of low ecliptic
latitude $\beta$ was also rejected, regardless of mask, in the disk region.
For $J$, $K$ and $L$, $|\beta|\!>\!15^\circ$; for $M$, $|\beta|\!>\!20^\circ$.

\subsection{The Optimization}

The maps were divided into bulge and disk regions. The bulge region is
defined as $|b|\!<\!12^\circ$, $|l|\!<\!20^\circ$; the disk region is 
everywhere else. The model was optimized separately for multiple samples of
pixels taken from the DIRBE maps. A sample of pixels was chosen, the fit made, 
its figures of merit and parameter values recorded, a new sample chosen, a new
starting point in parameter space chosen at random, and the process repeated. 
The disk region was represented by 660 pixels, only 60 taken from 
$|b|\!>\!20^\circ$, where a line of sight spans a smaller portion 
of the Galactic plane. The bulge region was represented by 170 pixels, out of 
the possible 6,795 which remain after the primary mask has been applied. The 
pixels comprising each sample were selected quasi-randomly, to ensure uniform 
spatial coverage without imposing the artificial regularity of a grid. In the 
disk region, the density of selected pixels was low enough that each sample was 
virtually unique, but in the bulge region some pixels were selected more than 
once, and some (5\%) never. The results presented are the 
averages over fits to 60 samples, but due to overlap between samples in the 
bulge region, this is no better than having only 37 independent samples of the
same size. The number of pixels in the bulge region limits how well the mean 
of any given parameter can be determined. 

In fitting the model, the figure of merit was calculated separately for each
band in each of the two regions, then weighted means taken to form a disk 
figure of merit and a bar figure of merit. In the disk region, $M$ is the least
trustworthy band due to residual zodiacal light, and $K$ and $L$ the most 
trustworthy. I assigned bands $J$ through $M$ the relative weights 
1.0, 2.0, 2.0, 0.1. In the bulge region, zodiacal light residue is negligible
and extinction makes $J$ the least useful band. There I assigned the relative 
weights 0.1, 2.0, 3.0, 3.0. Changing the relative weights in either region had
little effect on the final result, as long as most of the weight was not given
to either $J$ in the bulge region or $M$ in the disk region.
The figures of merit for the disk and bar were then averaged to arrive at a 
single figure. The relative weights of disk and bulge proved unimportant. Few 
of the parameters significantly affected the quality of fit in both regions. 

In Paper I, the figure of merit was not the $\chi^2$ but a statistically 
robust quantity:
$\psi\equiv<\!\!|\mbox{\em{DATA}}_j\!-\!
\mbox{\em{MODEL}}_j|/\sigma_j\!\!>,$
where $j$ denotes pixel number and the angular brackets a bisquare-weighted 
average over pixels. Unlike $\chi^2$, $\psi$ can accomodate small-scale 
irregularities in Galactic structure, such as the deep patchy extinction 
associated with molecular clouds, that remain even after applying the 
exclusion mask, and defy any smooth model. Another advantage of a robust 
figure of merit is that by giving less weight to a lesser component it aids in 
cleanly dividing the Galaxy into components: bar plus disk plus---whatever 
appears in the final residuals. Estimates of the measurement errors $\sigma_j$ 
were obtained from the scatter about polynomial fits to the local surface 
brightness. (See Paper I.) Although $\psi$ has the basic characteristics 
desired of the figure of merit, it makes no quantitative statement about the 
validity of the model in any case, and it has been replaced by a measure 
that is simpler and at least makes more intuitive sense: the mean absolute 
fractional deviation (MAD), obtained by substituting $\mbox{\em{DATA}}_j$ for 
$\sigma_j$ in the equation above. 

The optimization algorithm was based on the simulated annealing method 
(Press et al.\markcite{numrec} 1992). It never failed to converge.

\section{The Model}
My intention was to build a simple model. I still consider the model simple, 
overly simple in some respects, but, grudgingly, parameter by parameter, it 
grew to a minimum of 47 free parameters. Seven pertain to the data reduction, 
11 to the dust layer, 15 to the disk, 12 (to 14) to the bar, and two to the 
location of the Sun. A few of the symbols representing the parameters 
have been changed since Paper I in the interest of clarity, as their number 
has multiplied; these changes are noted in the text. The parameters are 
explained in the subsections that follow, and are labeled later on when their 
values are tabulated. 


Modeling was done in Galactocentric cylindrical coordinates ($R,\theta,Z$), 
with the Sun at $(R_0,180^\circ,Z_0)$, and $\theta$ measured counter-clockwise, 
looking down from positive $Z$. The parameter $Z_0$ is free.

The model consists of an exponential stellar disk, a bar, and an exponential
disk of diffuse dust. A thick disk is not included. As noted in Paper I, a 
thick disk is superfluous to reproducing the observed surface brightness. If 
one exists, it is too faint for the DIRBE to discern. After some 
experimentation with a halo model, a halo was rejected for the same reason.

The intensity at frequency $\nu$ in the direction $(l,b)$ is obtained 
by integrating the volume emissivity per unit solid angle $\rho$ along the 
line of sight $s$:
\begin{equation}
  I_\nu\,(l,b) = \delta_\nu +
                 \int_{(l,b)}\!\!\!ds \;(\rho_\nu^{\em Disk} + 
 \rho_\nu^{\em{dust}} + \rho_\nu^{\em Bar}) \; e^{-\tau_\nu(s)}. 
\end{equation}
The additive offset $\delta_\nu$ absorbs the extragalactic background and 
possibly zodiacal light and filtering artifacts. The extinction term 
$\exp{(-\tau_\nu(s))}$ is calculated from the dust model. 

\subsection{The Model of the Stellar Disk}

The modeled disk is exponential in $R$,  outer-truncated in $R$, 
inner-truncated in $R$, $\mbox{sech}^2$ in $Z$, and warped. The model does not 
force the truncations and warping, but allows them to occur if the data so 
dictate.

For $R\!\leq\!R_w$, another free parameter, the modeled disk is flat. 
For $R\!>\!R_w$, it is allowed to bend upward on one side and downward on the 
other, with a straight line of nodes at azimuth $\theta_w$. Letting 
$u\equiv R-R_w$, the mean elevation from the flat reference plane established 
by the inner Galaxy is: 
\begin{equation}
\overline{Z}(R,\theta)=\left [ c_1\:u+c_2\:u^2+c_3\:u^3
\right ] \sin(\theta\!-\!\theta_w). 
\end{equation}

In Paper I, giving the disk a hole in its center, as in a Freeman type II 
galaxy (Freeman\markcite{free} 1970), produced more plausible 
residuals than leaving it purely exponential. Holes have been found in the
disks of other barred spirals (Ohta, Hamabe \& Wakamatsu \markcite{oh}1990),
though not all (Baggett, Baggett \& Anderson \markcite{bag}1996), and this 
is a natural outcome in $N$-body simulations, in which a bar forms from an 
instability in the disk (Hohl\markcite{hohl} 1971; Schwartz\markcite{schwartz}
1984; Noguchi\markcite{nogu} 1996 and references therein). Therefore the 
modeled disk is permitted a central hole. Since the inner part of the disk 
cannot be expected to be axisymmetric in the presence of a strong bar, the 
hole is permitted an eccentricity. The hole is implemented through the 
function:
\begin{equation}
H(R,\theta) = 1-e^{-(R_H/O_R)^{O_N}},
\end{equation}
where $O_N$ is real and $R_H^2 = X'^2 + (\varepsilon Y')^2$ is calculated in 
the bar frame, in which $X'$ is defined by the bar's major axis. The axis ratio
of the hole is $\varepsilon$. This $H$ was the best of three three-parameter 
functions tried, and arose naturally from the best of several two-parameter 
functions, $H=1-\exp(-(R_H/O_H)^2)$. Freeman's \markcite{free} (1970) suggested 
$H=\exp(R_{min}^3/R^3)$ was not among the better two-parameter functions,
being too steep. The less-successful three-parameter functions were of logistic
and algebraic forms.

The full expression for the volume emissivity of the disk in each of the four 
bands is: 
\begin{equation}
\rho_\nu^{\em Disk}=\rho_\nu^{\em Disk}(0) \; H \; e^{-R/h_r} \; 
        \mbox{sech}^2(\frac{Z-\overline{Z}}{h_z}), 
\end{equation}
where 
\begin{equation}
h_r = \cases{\mbox{a free parameter}, & $R<R_{max}$;\cr
             0.5\mbox{ kpc},          & $R>R_{max}$.\cr} 
\end{equation}
The truncation radius $R_{max}$ and the central emissivity (in the absence
of a hole) $\rho^{\em Disk}_\nu(0)$ are also free parameters. (The symbol 
``$\rho(0)$" replaces the ``$k$" used in Paper I.)

The scale length $h_r$ and scale height $h_z$ are constant. This assumption
may not be strictly valid in the outer Galaxy, where dark matter comes to 
dominate the mass. In particular, $h_z$ may grow with $R$, as does the scale
height of the \ion{H}{1} layer (Burton\markcite{burt} 1976). I attempted to 
allow for this effect by including two additional parameters: a rate of change
of $h_z$ with $R$, and a radius beyond which the change occurs. This proved 
fruitless. Whether the scale height was a function of $R$ or $R^2$, the
additional freedom did not lead to a significant change in the figure of merit
of the fit. It did allow the truncation radius $R_{max}$ to be moved outward a
kiloparsec or two, but did not provide justification for the move, or for 
itself, and so was dropped.


\subsection{The Model of the Dust Layer}

The model of the dust layer is used to calculate extinction and emission by
dust, and possibly some of the light scattered by dust. The dust layer has the 
same form as the stellar disk, though with its own scale height ($h_z^d$), 
scale length ($h_r^d$), and hole parameters ($O_R^d$ and $O_N^d$). The 
eccentricity of the dust layer's hole is forced to that of the disk. The dust 
layer is given a warp with the same line of nodes, but an amplitude that 
differs by a scale factor $x^d$: $\overline{Z}^d=x^d\, \overline{Z}$. This was 
done because Freudenreich et al.\markcite{warplet} (1994) noted that, in 
surface brightness maps, the dust layer as observed in the far-infrared 
appears to be warped through a greater amplitude than the stellar disk, and 
because the mean distances of the disk stars and the dust that is important to 
the model (foreground dust) may not be the same. Two changes have been made 
since Paper I. The radial dependence of the dust density, in the absence of a 
central hole, has been changed from $0.5\:\mbox{sech}(R/h_r^d)$ to 
$\exp(R/h_r^d)$ to be consistent with the form of the stellar disk; and due to 
the greater sensitivity to extinction, now that more of the inner Galaxy is 
included, the extinction law of Rieke \& Lebofsky\markcite{rl} (1985) has been 
replaced by a power law in wavelength, $A_\lambda=A_J[\lambda/1.25]^{-\alpha}$,
where $A_J$ and $\alpha$ are free. 

It is likely the scale height of the dust layer, like that of the \ion{H}{1} 
layer, grows outside the solar circle, but the optical depth there is too low
for the model to grasp such detail.

Dust emission at these wavelengths is assumed to be due to fluorescence or
stochastic heating caused by {\em O} and {\em B} stars with the same scale 
length as the old disk, but with the warping ($x^d\,\overline{Z}$) and the
scale height ($h_z^d$) of the dust. The dust emissivity at any point is 
proportional to the product of the dust density and the UV emissivity within 
a UV absorption length ($\propto \left[\mbox{dust density}\right]^{-1/3}$), 
so that 
\begin{equation}
\rho_\nu^{\em dust}= \rho_\nu^{\em dust}(0) \:H^d\:H\:
e^{-R/h_r \,-2R/3h_r^d}\;   
\mbox{sech}^{10/3} (\frac{|Z-x^d \, \overline{Z}|}{h_z^d}).
\end{equation}
Note that as a product of the dust density and a stellar emissivity, the 
emissivity of the dust falls off more rapidly in $R$ and $Z$ than either alone.

Dust-scattered starlight is neglected. However, since it is proportional to 
the dust density, and nearly isotropic at these wavelengths, there will be an 
apparent correlation between scattering and emission, and some scattered light
will find its way into $\rho_\nu^{\em dust}$.

\subsection{The Bar Models}

Other spiral galaxies show a great variety in the shapes of their bars, as 
they do in many things, but some generalities have been drawn. A typical bar is 
straight. Its brightness along the major axis varies from roughly exponential 
to roughly flat, and often ends abruptly. The flatter bars tend to belong to 
galaxies of earlier Hubble type, and to be stronger, often joining directly to 
spiral arms or a ring (Elmegreen \& Elmegreen\markcite{ee85} 1984; Elmegreen 
et al.\markcite{ee96} 1996). The outer part of a bar often has a somewhat 
rectangular appearance when viewed face-on (Ohta et al.\markcite{oh} 1990). 
See Kormendy \markcite{korbarsum} (1977) and Sellwood \& 
Wilkinson\markcite{sellbarsum} (1993) for general information.

The simplest bar shape is an ellipsoid. More realistically, a ``generalized
ellipse" has been proposed by Athanassoula et al.\markcite{athshape1} (1990) 
for the two-dimensional case of a galaxy observed face-on:
\begin{equation}
R_s^C = \left( \frac{|X'|}{a_x} \right)^C +
        \left( \frac{|Y'|}{a_y} \right)^C,
\end{equation}
elliptical when $C=2$, diamond- or lozenge-shaped when $C\!<\!2$, and boxy 
when $C\!>\!2$. This can be generalized to three dimensions:
{ \everymath{\displaystyle}
\begin{eqnarray}
R_{\perp}^{C_{\perp}} = & \left( \frac{|X'|}{a_x} \right)^{C_{\perp}}+
            \left( \frac{|Y'|}{a_y} \right)^{C_{\perp}},  \\
\rule{0in}{3ex}
R_s^{C_{\parallel}} = & R_{\perp}^{C_{\parallel}}  +
            \left( \frac{|Z'|}{a_z} \right)^{C_{\parallel}}.
\end{eqnarray}}
The effective radius is $R_s$; the scale lengths are $a_x$, $a_y$, and $a_z$;
and $C_{\perp}$ and $C_{\parallel}$ are the face-on and edge-on shape 
parameters. A bar may appear diamond-shaped from one vantage point and boxy 
from another. Athanassoula et al.\markcite{athshape1} (1990) and Athanassoula 
\markcite{athshape2} (1992a) found that both the ellipticity and the shape
parameter of face-on galaxies varied with $R$, but after some unsuccessful
experimentation with an $R$-dependent $C_{\perp}$ and $a_x/a_y$ ratio, I
decided to leave these parameters single-valued.

The radial dependence of the bar must cover the range of flat to exponential.
I have tried three functions flexible enough to model this characteristic, 
when coupled with a term that truncates them radially:
{ \everymath{\displaystyle}
\begin{eqnarray}
\mbox{Model S}: & \rho\propto &\mbox{sech}^2(R_s) \\
\mbox{Model E}: & \rho\propto &\exp(R_s^{-n}) \\
\mbox{Model P}: & \rho\propto &(1+(R_s/R_c)^n)^{-1} 
\end{eqnarray} }
Model S has the fewest parameters. At first, the power of the sech function 
was left free, but in test runs it settled so close to a value of 2.0 that I 
fixed it there. Giving the bar and disk the same functional $Z$-dependence has 
the advantage of allowing the direct comparison of scale heights. Model E, an 
exponential-to-a-power, fit almost as well, but at the price of an extra free 
parameter, the power $n$. An $n$ of 1.0 would give us an exponential bar, the 
form recommended by Stanek et al. \markcite{sta}(1997). An $n$ of 2.0 would 
match the radial dependence of the best-fitting model of Dwek et 
al.\markcite{dwek} (1995). Model P, a power law with core radius $R_c$, led to 
a distinctly inferior fit. 

The bar would possibly be better fit using a different function, rather than 
just a different scale length, along each axis, but I am reluctant to enter 
this wilderness of functional combinations without the firm guidance of theory.

To truncate the bar at radius $R_{end}$, its density is multiplied by a 
Gaussian function with scale length $h_{end}$. Both $R_{end}$ and $h_{end}$
are free parameters. In model S, for example, the volume emissivity of the 
bar is: 
{\everymath{\displaystyle}
\begin{eqnarray}
\rho_\nu^{\em Bar}=\rho_\nu^{\em Bar}(0)\;\mbox{sech}^2(R_s), & & 
  R \leq R_{end}; \\
\rho_\nu^{\em Bar}=\rho_\nu^{\em Bar}(0)\; \mbox{sech}^2(R_s) &
 e^{-(R_s\!-\!R_{end})^2/h_{end}^2}, & R>R_{end}. 
\end{eqnarray}  }

The bar has two more degrees of freedom: a tilt angle $\theta_0$ (a clockwise 
rotation about the $Z$ axis from the Sun-center line) and a pitch angle (the 
angle between the bar's major axis and the Galactic plane). 

A dust lane often runs along the leading edge of a bar, but is not 
included in my model, in the belief that masking the low-latitude sky would 
hide it. If a dust lane is present, we must search for it in the residuals.

\section{The Fit and Its Residuals}

The mean absolute deviations of the models, obtained using the primary mask, 
are given by Table 1. These are averages over the 60 fits. The MAD was the
quantity optimized, but $\psi$ and the $\chi^2$ per degree of freedom were 
also calculated and are included. Ideally, the values of $\chi^2$ and $\psi$ 
would be 1.0 and 0.77 respectively, but due to the presence of the young disk,
the figures of merit would not approach the ideal values even if the models of
the bar and the old disk were perfect. In any case, the figures of merit are 
only relative and do not provide confidence limits on a model. The formal 
uncertainties are omitted, except in the last column, containing the weighted 
mean values. As expected, by all measures the $J$-band fit is the worst in the
bar region, and the $M$-band fit the worst in the disk. The $L$ band, with low 
extinction and very little remaining zodiacal light, provides the best fit, 
judging by $\psi$ and $\chi^2$. (The MAD cannot be used to compare different
bands.) 

In the disk region, the figures of merit of all the models are approximately
equal, indicating the exponential part of the disk has been successfully
decoupled from the inner part. In the bar region, Model S has a small but 
consistent superiority over Model E, while Model P is a poor third. When 
the minimal mask was used, all figures of merit worsened, MAD by $\sim10$\% 
over its previous value, but the relative quality of fit among the models 
did not change. (They were not tabulated for this reason.)

A penalty incurred by smoothing away point sources must be discussed before 
contour plots of the data and model are presented. Because of the broad DIRBE 
beam, it is difficult distinguishing between nearby stars and pointlike but
extended distant sources, such as those on the brightness crest of the inner 
Galaxy. As a consequence, the brightness crest is rounded off. The effect of 
this on the old disk can be gauged by applying the same procedure to the 
modeled $L$-band surface brightness map. In Fig.~\ref{smoothfac} we have the 
difference between the unsmoothed and smoothed maps, from Model S, as a 
fraction of the unsmoothed map's surface brightness. The effect is greatest in 
a narrow strip approximately $2^\circ$ wide, the size of the minimal mask. An 
``X" appears where the corners of the bulge have been eroded, but it is very 
faint and fills only a small part of the bulge region. This figure tells us 
that if data and model are to be compared over the whole sky, and not just 
over the non-masked area, both must be either smoothed or unsmoothed. (It also 
reminds us that a boxy bulge may appear X-shaped if a non-boxy model is 
subtracted from it.)

Fig.~\ref{contour90} compares the smoothed DIRBE $L$-band map to the smoothed
map made using Model S, with the primary rejection mask overlaid. In this
two-dimensional projection, Models E and S are very similar. In the 
non-masked region the agreement between data and model is excellent. The major 
discrepancies occur in the direction of the local spiral arm or spur in Cygnus,
and toward the molecular clouds in Ophiuchus, Orion, and Taurus.

While Fig.~\ref{contour90} tells us how well model and data agree, the 
disagreements are best examined by focusing on the residuals. In 
Figs.~\ref{bartours1} through \ref{bartours4}, the smoothed modeled 
surface brightness, using Model S and the primary mask, has been subtracted 
in stages from the smoothed DIRBE $J$ through $M$ sky maps. In the $J$ map, 
the final residuals are predominantly negative. There is more extinction, 
especially at positive latitudes, than the model predicts. Proceeding from 
the $J$ through the $K$, $L$, and $M$ bands, the extinction lessens and the 
negative residue approaches zero. It does not reach zero, however, which
suggests a flaw in the model, perhaps in its treatment of the dust layer.

There is much information in these plots of residuals, but as that information
does not include distances, they should be interpreted with care. Most 
prominent in the residuals is a bright nucleus roughly $2^\circ$
(or 300 pc, at a distance of 8.5 kpc) in diameter. This becomes brighter
and morphologically simpler as the observing wavelength increases, and probably 
does occupy the center of the Galaxy, a unique environment in which the model
is certainly inadequate (see, for example, Morris \& 
Serabyn\markcite{mor} 1996). There is also a bright narrow ridge in the inner 
Galaxy along the Galactic midplane, probably created by stars associated with 
the young disk, red supergiants in particular. The ridge is brighter at 
positive longitudes, fainter and partially broken at negative longitudes. It is 
tempting to link this asymmetric feature to the bar. Hammersley\markcite{ham} 
et al. (1994) identified peaks in the DIRBE $K$-band surface brightness at 
$l=21^\circ,\;27^\circ$ as probably originating in star-forming regions at the
near end of a bar, which would need to be long ($3.7-4$ kpc) and viewed almost 
broadside (at an angle of $75^\circ\pm5^\circ$). Calbet\markcite{cal} et al. 
(1996) developed this idea further. Since the oblong boxy bulge seen in the 
DIRBE maps is much too narrow to be explained by such a bar, the bar must be 
very thin in $Z$, and exist in addition to what they refer to as the ``thin 
bulge." It appears, however, that the ``thin bulge" is actually a strong 
bar stretching more than three kpc from the Galactic center, and two bars of
roughly the same length are not observed in other spiral galaxies. These peaks
of brightness noted by Hammersley et al. (1994) do occur in my 
model-subtracted maps. They appear even brighter and extend further in $l$ in 
the $L$ and $M$ bands. Yet I prefer to locate these peaks on an arc or arm 
trailing the bar's near end, or on a bright segment of a bar-circling ring.
I agree with Calbet\markcite{cal} et al. (1996) that extinction caused by a 
dust lane could cause the brightness ridge to appear fainter in the fourth 
Galactic quadrant. In the first quadrant we probably see the trailing edge of 
the near end of the bar and in the fourth the leading edge of the far end. The 
dips in the surface brightness for $-12\!>\!l>\!0^\circ$ could well be 
due to dust within or at the leading edge of the bar. At more negative 
longitudes I believe we must search for other explanations.

When the minimal mask is used there is no great change in any one parameter.
Extinction toward the inner Galaxy decreases slightly, and the residuals
acquire a small shift toward the negative, as can be seen in 
Fig.~\ref{bartour3_0}, the final $L$-band residuals of Model S. 
The surface brightness toward the Galactic center and the brightness crest
to either side simply cannot be accounted for by the model. Using the minimal 
mask probably gains us nothing but the greater opportunity of being biased by
features extraneous to the bar and old disk. Fig.~\ref{bartour3_0} is the only
figure presented that involves the minimal mask, but the values of the model 
parameters will be tabulated for both masks (Tables 2 and 3, discussed in 
the next section), to give some idea of the possible bias entailed.

Now we move to the unsmoothed maps. Fig.~\ref{strips} displays pseudo-color 
images from the $J$, $K$, and $L$ bands before and after model subtraction. 
Translating from color to surface brightness may be difficult because of the 
unusual color table, which is better at displaying shape and detail, but that 
information is available in the contour plots already discussed. 
Fig.~\ref{coloresid} combines the $J$, $K$, and $L$ residual maps in a 
three-color image. The aforementioned Galactic nucleus and brightness ridge 
are evident. Extended white patches mark the locations of low-extinction 
windows. Appearing as pink or red are directions in which the extinction is 
unusually heavy or there is significant emission by dust or luminous 
dust-shrouded stars. There are spurs of heavier extinction reaching upward 
from $l\approx0^\circ$ and $l\approx\pm25^\circ$. These have counterparts in 
far-infrared image presented in Paper I (Fig. 2b) and in maps of CO emission 
(Dame et al. 1987, Fig. 6), which is strongly correlated with dust density. 

Fig.~\ref{cobemap} presents the same composite, before and after model 
subtraction, on a logarithmic brightness scale, so the bulge does not 
overwhelm the disk. The fact that the disk and the bulge are morphologically 
identical at the three wavelengths, and have been scaled to the same 
emissivity, causes the three colors to sum to a gray haze where the column 
density of dust is low; the residuals actually appear brighter than the 
original DIRBE image.

\section{Discussion of the Models}    

\subsection{The Tabulated Parameter Values}

The values of the parameters of the three models were normally distributed.
Their means and the standard deviation of the means are listed in Tables 2 and 
3. Of course, the standard deviations tell us more about the models' 
consistency than about their validity. Comparison of Models S and E, and of 
the results obtained using the primary and the minimal masks, probably provide
a better idea of the true uncertainties. In going from the primary to the 
minimal mask, the major change is a decrease in extinction.

It might have been misleading to include the coefficients $c_i$ of the cubic 
polynomial that describes the amplitude of the warp, since the 
coefficients are not independent. Instead, the shape of the warp is shown 
graphically in Fig.~\ref{warps}. As noted in Paper I, the orientation and 
magnitude of the warping is consistent with that of the \ion{H}{1} layer the 
shape of which, however, is not known very precisely for $R\!<\!12$ kpc.

In all models, the deletion distances $D_\nu$ are $\approx$ 470, 520, 560,
and 560 pc for bands $J$ through $M$, and the offsets in surface brightness
$\delta_\nu$ are 84, 64, 31, and 14 kJy sr$^{-1}$. These numbers include the
extragalactic background, but are also functions of the data reduction process, 
so too much should not be read into them. 

The tilt angle of the bar, $\theta_0$, is grouped with $R_0$ and $Z_0$. To
place it with the bar parameters would be chauvinistic. 

Since they differ only in bar model, it is not surprising that the models 
agree on most disk and dust-layer parameters. The plausibility of the 
parameters describing the stellar disk was discussed in Paper I, but the 
parameters describing the dust layer require further discussion.

\subsection{Comparison to Paper I, and a Correction}

Paper I contained two errors, both concerning the dust layer. One occurs only 
in the values of $A_J$ found in Tables 2 and 3 of that paper. They have been 
erroneously multiplied by a factor of two. For example, for $R_0=8.5$ kpc, 
$A_J$ should read ``0.213" mag kpc$^{-1}$. This error is found only in the 
tables. It is also moot, because of an error in the software. The extinction 
in the different bands was to be related by the law of Rieke 
\& Lebofsky (1985), which has a power law index $\alpha=1.765$ in bands $J$ 
through $M$. In the modeling program, this index became 1.33, which affected 
other parameters of the dust layer model. What should those parameters have 
been? Using Model S without a disk-hole, and minimizing $\psi$, as in Paper 
I, rather than the MAD, led to stellar disk parameters almost identical to 
those in Paper I. For $R_0=8.5$ kpc, $h_r=2.63$ kpc, $h_z=0.336$ kpc, 
$R_{max}=12.03$ kpc, $Z_0=15.60$ pc, and so on. The dust 
layer parameters did change: $h_r^d=3.13$ kpc, $h_z^d=0.20$ kpc, $A_J=0.132$ 
mag kpc$^{-1}$, and $\alpha=1.79$. Of these new numbers, the newer, smaller,
value of $A_J$ is closer to the expected value. The value of $h_z^d$ is 
larger than expected, even when the inner Galaxy is include and $h_z^d$
drops to 0.15 kpc, but here my expectations were probably at fault. A review 
of \ion{H}{1} in the Galaxy by Dickey\markcite{dic} \& Lockman (1990) models 
the vertical structure of the \ion{H}{1} layer, $0.4\,R_0\!<R\!<\!R_0$, using 
the sum of two Gaussian terms and one exponential. This is shown in 
Fig.~\ref{zdust}, along with the profile derived from Model S, 
$\rho^{dust}\propto \mbox{sech}^2(Z/0.152\mbox{ kpc})$. For $Z\!<\!200$ pc, 
the two curves almost coincide, but at higher elevations the density falls off
more rapidly in my model. If the \ion{H}{1} and dust layers have the same 
vertical structure, and Dickey \& Lockman (1990) have accurately described it, 
then the model underestimates the extinction at higher elevations. Could this 
have caused the areas of negative residuals that lie a few degrees off the 
brightness crest of the inner Galaxy? Substituting the $Z$ dependence of the 
\ion{H}{1} for the sech$^2(Z)$ term in the dust layer produced as good a fit 
to the data, with the parameter values almost unchanged, but did not improve 
the residuals. It is still possible, though, that a more sophisticated 
treatment of the dust layer is called for.

One other change since Paper I is the switch from a dust density 
$\propto\,0.5\;\mbox{sech}(R)$ to a simple exponential. This did not make a 
significant difference.

In all three models, the stellar disk parameters are consistent with those
of Paper I, with the exception of $Z_0$. In all models, the distance of the 
Sun from the Galactic plane rises several tenths of a parsec to a parsec
when the disk is allowed a central hole. 

\subsection{The Luminosity of the Dust Layer}

The central emissivities of the dust layer, $\rho_\nu^{dust}(0)$, are of the 
same order of magnitude as those of the disk and bar, but then, there is no 
dust in the center of the Galaxy, according to the model, and at larger $R$ 
the more rapid radial decrease in the emissivity of the dust causes it to drop
well below the stellar emissivity of even the $M$ band. According to Model S, 
if we observed the Galaxy face-on, we would see the ratio of $L$-band dust/disk 
surface brightness peak at $R=2.74$ kpc, at a value of 4.1\%. At $R_0$ the 
ratio would be only 0.5\%. In the $M$ band, these ratios would be 12\% and 
1.5\% respectively. These numbers would undoubtedly be higher if dust 
associated with the masked-out molecular clouds were included, but 
near-infrared dust emission would still be a minor component of the Galaxy's 
luminosity. The radius of the dust layer's hole is smaller than that of the 
disk. The presence of the three-kiloparsec molecular cloud ring and possibly 
a stellar ring, both absent in the model, and the model's lack of sensitivity 
to dust in the background of most of the stellar emission, suggest this result 
be treated with caution. 

\subsection{The Disk and its Hole}

The parameters that describe the exponential part of the disk (approximately
$R>5$ kpc) are almost unchanged from Paper I. The radius of the Galactic disk 
is still a mere 12 kpc. Robin, Cr{\'e}z{\'e}, \& Mohan\markcite{rob} 
(1996) placed the edge of the disk 5.5 kpc from the solar circle, or 
$R\approx14$ kpc. Their distance measure, based on $V-I$ color, may be wrong
or, as previously discussed, one or more of the basic assumptions of my 
model may break down in the outer Galaxy.

The central hole in the disk appears in all three models. Forcing the disk
to be exponential all the way to its center results in poor fits in all models.
The values of the MAD in the bar region goes from 4.2\% to 5.4\%, and the 
and the $\chi^2$ increases by a factor of 5. The exponential part of the disk 
does not change appreciably. It becomes slightly thinner: $h_r/h_z \approx 
7.8$, instead of 7.5. 

The axis ratio of the disk-hole is near the middle of the range of 0.7 to 1.0
that Buta\markcite{buta} (1986) gives for stellar rings circumscribing bars.
Strongly barred spiral galaxies that do not have rings are usually of grand 
design, with one arm trailing from each end of the bar. When ringed, 
they are usually multi-armed, with arms beginning at points on the ring that 
seem unrelated to the bar's orientation. The best evidence to date suggests 
the Galaxy is of the second type (Val{\'e}e\markcite{va} 1995). I suspect it
is also ringed.

\subsection{The Bar and the Disk-Hole}

In all three models the bar ends at the inner edge the Galaxy's molecular 
cloud ring, at $R\approx3.5$ kpc, in agreement with observations of other 
spiral galaxies, and with simulations that show a strong rotating bar sweeps 
up the gas and dust in its vicinity (Athanassoula\markcite{athdust} 1992b).
There seems to be a current consensus that rings form at the inner second 
harmonic resonance (Schwartz\markcite{schwartz} 1984), just inside the 
corotation radius, beyond which bars cannot extend (Contopoulos et al.
\markcite{con}1989). 

Fig.~\ref{faces} shows the face-on surface brightness predicted by the models 
(without dust). The bar of Model P seems the least realistic. It is a hybrid 
of power-law and Gaussian along its major axis. The index of the power law is 
5.0, higher than the values $\sim3-4$ often used in models of the Galactic 
halo, but not much higher than the best power-law fits of Stanek et 
al.\markcite{sta} (1997), in which $\rho \sim R^{-4}$. The bar/disk luminosity 
fraction is 0.56, and while estimates of this quantity vary greatly with 
galaxy and with measurement technique, the typical value for an early-type 
galaxy seems to be less than half this (Sellwood \& 
Wilkinson\markcite{sellbarsum} 1993). The appearance of the bar is also 
unusual in that its outer isophotes lie parallel to the $X'$ axis for most of 
the bar's length, but come to a point at its ends, rather than forming a 
blunt, boxy terminus.

The outermost contours of the Model E bar are not as pointed as those of 
Model P, but not as rectangular as one might expect. The bar/disk luminosity 
ratio is a plausible 0.33. The power of the exponential, 1.44, is intermediate 
between those of the best models of Dwek et al.\markcite{dwek} (1995) and 
Stanek et al. \markcite{sta}(1997). This is my second-best model, on the basis
of figure of merit and my subjective evaluation of its appearance.

My preferred model is Model S. Its bar is slightly shorter than the Model E 
bar, and tilted through a slightly greater angle. Its bar/disk luminosity 
ratio is 0.33, and it outer isophotes are clearly rectangular. The isophotes 
in its interior are diamond-shaped, with $C_\perp=1.57$. This contrasts with 
the findings of Athanassoula et al.\markcite{athshape1} (1990) that 
$C_\perp=2-4$ in the bars of SB0 galaxies. They warn that an apparent 
$C_\perp\!<\!2.0$ might result from the unintentional inclusion of a nuclear 
component that projects onto a circular area, or onto an elliptical area with 
long axis normal to the bar. To test for this, I excluded pixels within first 
$5^\circ$, then $6^\circ$, then $7^\circ$ of the Galactic center, refitting 
Model S each time. When the exclusion radius increased, there was no 
significant change in $C_\perp$, so contamination by a nuclear component is 
probably not important. From the residual plots already shown, such a 
component may well exist, but have a much smaller scale height than the bar.

Diamond-shaped isophotes are not uncommon in bars,
however. In the simulations of Contopoulos et al.\markcite{con} (1989), 
orbits near the center of a bar are elliptical (the $x_1$ family of orbits), 
farther out they are increasingly diamond-shaped, and near the ends, 
rectangular (the 4:1 family of orbits). Quillen, Frogel \& 
Gonzalez\markcite{quil} (1994) also note a transition from diamond-shaped 
to rectangular orbits in NGC 4314, and suggest it occurs at the $m=4$ inner 
Lindblad resonance. Whether or not this is the case, the contour plot of 
Model S in Fig.~\ref{faces} resembles that of NGC 4314 in Quillen et 
al.\markcite{quil} (1994). Athanassoula\markcite{ath86} (1996) 
wondered if bars would appear as boxy in the near-infrared as in the $B$ band.
In the case of our Galaxy, the answer would seem to be, ``Not quite."

With a $C_\parallel$ of 3.5, the bar is definitely boxy when seen edge-on.
In the best bar model of Dwek et al.\markcite{dwek} (1995), 
$C_\parallel \equiv 4.0$. In this model (``G2"), $C_\perp \equiv 2.0$, the 
bar power $n \equiv 2.0$, and the bar cutoff is fixed at either 2.4 kpc (from 
Binney et al.\markcite{bin} 1991) or 5.0 kpc (from Weinberg 
\markcite{wein}1992), though with a scale length $a_x \sim1.7$ kpc, any cutoff
beyond 3.5 kpc is probably moot. 
Instead of modeling the disk, they extrapolated its surface brightness in the 
DIRBE maps inward from larger longitudes to the region of the bulge. (Since
the extrapolation did not take into account a central hole in the disk, this
probably produced an overestimate.) To further minimize the effects of the disk,
they chose a relatively small bulge region: $3^\circ\!<\!|b|\!<\!10^\circ$,
$|l|\!<10^\circ$. The $K$, $L$, and $M$ bands were fitted individually, with
the $\chi^2$ as the figure of merit. The agreement between our results is 
fairly good, despite the differences in method. Taking their best (and most
consistent) model, G2 with a 5 kpc cutoff, and ignoring the $K$ band, which 
differs greatly in scale lengths from the other two, we find a tilt angle of 
$9.5^\circ$ and axis ratios of $1.75:0.62:0.42$ kpc. The tilt angle and the
ratio of scale lengths agree with those of my models. However, other Dwek 
models which have $\chi^2$'s nearly as good as G2 lead to very dissimilar bars. 
I believe fewer and more general models would have produced more consistent
and realistic bars. Some (such as Kuijken\markcite{kuik} 1996) have commented 
on the inherent limitations in using surface photometry to reconstruct a 
three-dimensional bar. The problem is admittedly challenging, but we should 
not underestimate the amount of information offered by the DIRBE maps; even
a 47-parameter model may be far from exhausting it.

The Stanek results are more consistent than those of Dwek, at least in 
tilt angle. They applied the Dwek models and several of their own to the 
distribution of ``red clump" giants, which have a narrow range of intrinsic
luminosity, in 12 fields toward the bulge. They assumed $R_0\equiv8.0$ kpc. 
All the Stanek models returned a tilt angle $\sim20^\circ$ ($14^\circ$ to 
$34^\circ$). Their best models were power-law and exponential. The power-law 
models (exemplified by model ``P2": $\rho\propto [R\,(1+R)]^{-2}$) had axis 
ratios of $1.10:0.45:0.29$ kpc. The exponential models (``E2": $\rho \propto 
\exp(-R)$) had axis ratios of $0.94:0.34:0.26$ kpc. Their Gaussian model 
similar to the Dwek G2 model had axis ratios of $1.33:0.56:0.45$ kpc. Their 
coverage of the area of the bulge is still sparse; only six distinct 
directions were sampled, and only two outside the latitude strip 
$-5^\circ<b<-3^\circ$.                         

The bar postulated by Binney et al.\markcite{bin} (1991) is tilted only 
a degree or two more from the Sun-center line than my Model S bar, but 
is significantly smaller and weaker. The bar is of the power-law type,
ending within a corotation radius fixed at 2.4 kpc. It is unclear how 
the existence of a bar like that of Model S would affect their interpretation 
of gas orbits in the inner Galaxy. 

Profiles of the disk and bar along the major and minor axes of the bar are 
also shown in Fig.~\ref{faces}, along with profiles obtained by averaging 
disk and bar over azimuth angle. In S and E the averaged profile remains 
approximately exponential as it continues inward in $R$. Ohta et al. 
\markcite{oh}(1990) found this to be true in their sample of barred spiral 
galaxies, and cited it as evidence that the bar formed from an instability of 
the disk, with little redistribution of stars in the radial direction. The 
similarity in near-infrared color of bar and disk also supports this theory. 
(If we normalize the central emissivities to their $L$-band values, we get 
2.33, 1.89, 1.0, 0.51 for $J$ through $M$ in the disk and 2.26, 1.93, 1.0, 
0.50 in the bar.) The greater scale height of the bar, 0.43 kpc 
vs 0.34 kpc for the disk, is consistent with Noguchi's\markcite{nogu} (1996) 
theory that early-type bars form from an already-mature, thickened, Galactic 
disk.

The type I disk models are shown in Fig.~\ref{facenoho}. When the disk is not 
permitted a hole, the bar becomes less luminous (20\% of the disk luminosity) 
and very diamond-shaped ($C_\perp=1.28$). Neither the face-on views nor 
the profiles seem credible. 

\subsection{The Parameter $R_0$}

All models showed a slight sensitivity to $R_0$ in their figures of merit, but
only in the disk region. This can probably be attributed to extraneous factors,
namely zodiacal light residue or an artifact of its removal; or to large-angle
features of the young disk. To see how the values of the model parameters 
depend on our distance from the Galactic center, I repeated Model S for values
of $R_0$ spanning the range of 7.5 to 9.5 kpc, approximately the range found 
in the current literature. (See Reid 1993 for a review.) The results are 
listed in Table 4. The formal errors have been omitted, but they are similar 
to those given by Table 2 for $R_0=8.5$ kpc. 

Most of the parameters are dependent on $R_0$ in a simple way, and it might
be possible to determine $R_0$ if one of these parameters were known with 
confidence through some other means, but at present they are at least as
uncertain as $R_0$. One parameter which might provide a lower limit is the
extinction at the Sun's position. At $R_0=7.5$ kpc, the local $J-$band 
extinction is only .066 mag/kpc, and if the power law of extinction were 
extended to the $V$ band, it would lead to an extinction of 0.3 $V$ mag/kpc, 
half what is considered a reasonable value. How significant is this 
discrepancy? One could argue that the term ``local" is open to interpretation, 
when dealing with something as patchy as extinction, or that, by masking 
directions with high optical depths, I may have introduced a bias toward 
underestimating $A_J$, but I think it very probable that $R_0>7.5$ kpc. 
No such discrepancy is obvious at the high end, $R_0=9.5$ kpc, though for this 
distance the bar becomes uncomfortably diamond-shaped, with a face-on shape 
parameter $C_{\perp}=1.48$. 

\subsection{Luminosities}

In Paper I, I fitted a Planck curve to the central emissivities of the disk, 
deriving an effective temperature of 3800 K and using this to estimate the 
total luminosity of the old stellar disk. Since the colors are virtually 
unchanged, and the same for bar and disk, the same is done here for disk+bar 
(according to Model S). Weighted by a 3800 K black-body spectrum, the 
effective frequencies of the DIRBE $J$, $K$, $L$  and $M$ bands are, 
respectively, 2.38, 1.36, 0.864, 0.615 $\times10^{14}$ Hz. The effective 
widths of the bandpasses are 5.85, 2.22, 2.18, 0.828 $\times10^{13}$ Hz. For
$R_0=8.5$ kpc, the luminosities are 39.2, 12.2, 6.3, 1.2 
$\times10^8 \; L_\odot$. Calculating the bolometric corrections for the four 
bands, and averaging them using the weights in Table 1, we obtain the 
luminosities of the disk and bar. Their sum is given is the last line of 
Table 4; for $R_0=8.5$ kpc, the total luminosity is 
$2.3\times10^{10}\;L_\odot$. There is little change when switching to Model E 
or the minimal mask; the total luminosity varies by less than 4\% among the 
two bar models and two masks.

\section{Conclusions}

The old stellar disk is approximated well by a function that is exponential
in $R$ and sech$^2$ in $Z$. It is warped similarly to the \ion{H}{1} layer,
has an outer edge and a central hole. The stellar emissivity peaks at 
approximately the inner radius of the ``three-kpc ring" of molecular clouds, 
and is truncated approximately four kiloparsecs beyond the solar circle.
The dust layer also has a central hole, slightly smaller but sharper than 
that of the disk. If $R_0=8.5$ kpc, the scale length of the disk is 2.60
kpc and the scale height 0.34 kpc, and the Sun is located 16.5 pc above its
midplane.

I have modeled the Galaxy's bar using three types of function: a power law 
with a core, an exponential-to-a-power, and a sech$^2$. Of these, the 
sech$^2$ function provided the best fit, but all three bar models agree on
several points:

The bar is strong and truncated at approximately the radius of the hole.
It has the same color as the disk, but a larger scale height. The bar lies
in the plane of the Galaxy and is tilted $9^\circ-15^\circ$ from the line 
between the Sun and the Galactic Center. There is probably a nuclear component 
with scale length $\sim100$ pc in addition to the disk, bar, and features 
attributable to the young disk, such as giant molecular clouds, spiral arms,
and possibly a circum-bar ring. 

Model S, the ``sech$^2$" bar, is the best model in terms of figure of merit, 
simplicity, and similarity to other barred spiral galaxies. According to this
model, if $R_0=8.5$ kpc, the bar has axis ratios of $1.70:0.64:0.44$ kpc. It 
is tilted $14^\circ$, and is one-third as luminous (in the near-infrared) as 
the disk. Other characteristics of the bar, disk, and dust layer are presented 
in Table 4.

How much faith can be placed in these results?

The most distinguishing feature of my model is probably its versatility.
It allows the Galactic disk to warp and to have an inner hole and a sharp
outer edge. It allows the bar to have its own truncation and a different 
``boxiness" when viewed from above or from the plane of the disk. 
Nevertheless, the model is not infinitely flexible, nor provably the best. 
Yet I find the similarity in the end products of the three bar models 
encouraging. Despite the constraints of their different functional forms, they 
seem to be converging on a single destination. There is no reason to believe 
that even the best of them has arrived at that destination, but its 
self-consistency, plausible parameter values and small, mostly explainable, 
residuals argue that it has come close. 

\acknowledgements
COBE is supported by NASA's Astrophysics Division. Goddard Space Flight Center 
(GSFC), under the scientific guidance of the COBE Science Working Group, was 
responsible for the development and operation of COBE. The author is grateful
to the COBE team and wishes to express special thanks to the surviving
DIRBEans, Rick Arendt, Nils Odegard, and Janet Weiland. And my thanks to the
anonymous referee.

\clearpage

\vspace{1in}
\begin{figure}
\caption{One sample of pixels selected for fitting the model. The area 
within the inner contours is rejected for being anomalously red. The area
within the outer contours is rejected from only the $J$ band. The dotted 
curves are at $\beta=\pm15^\circ$, within which zodiacal light residue may 
be significant.}
\label{sample}
\end{figure}

\vspace{1in}
\begin{figure}
\caption{The fractional increase in surface brightness (Model S) when the 
point-source-removal process is not applied. The outermost contour level is
the 0.25\%. Each level inward is a factor of two higher. The thick contour 
represents the primary rejection mask.}
\label{smoothfac}
\end{figure}

\vspace{1in}
\begin{figure}
\caption{The $L$ band (3.5 $\mu$m) DIRBE and Model S (smooth isophotes)
surface brightness over the full sky. The area within the dark contour was 
excluded when the model was fit to the data. The surface brightness due to 
point sources, as calculated by the model, was added to both. The contour
levels are: 0.09, 0.13, 0.18, 0.26, 0.52, 0.73, 1.04, 1.47, 2.08, 2.94,
4.17, 5.90, 8.36 MJy sr$^-1$.}
\label{contour90}
\end{figure}

\vspace{1in}
\begin{figure}
\caption{The $J$ band (1.25 $\mu$m) map (top), and after modeled disk (middle)
and modeled disk and bar (bottom) subtracted. The gray contours represent 
$I_\nu<0$. Contour levels are numbered in hexadecimal, starting at 0. The 
levels (in MJy sr$^{-1}$) are 0.06, 0.21, 0.39, 0.62, 0.91, 1.27, 1.73, 2.31,
3.04, 3.96, 5.11 (top); $\pm$ 0.08, 0.30, 0.59, 1.00, 1.54, 2.30, 3.32, 4.73,
6.65, 9.28 (middle); $\pm$ 0.07, 0.25, 0.49, 0.80, 1.21, 1.75, 1.75, 2.45,
3.38 (bottom).}
\label{bartours1}
\end{figure}

\vspace{1in}
\begin{figure}
\caption{The $K$ band (2.2 $\mu$m) map (top), and after modeled disk (middle)
and modeled disk and bar (bottom) subtracted. The gray contours represent 
$I_\nu<0$. Contour levels are numbered in hexadecimal, starting at 0. The 
levels (in MJy sr$^{-1}$) are 0.07, 0.23, 0.44, 0.71, 1.06, 1.51, 2.10, 2.84,
3.81, 5.05, 6.65, 8.71, 11.4, 14.8 (top); $\pm$ 0.09, 0.34, 0.68, 01.17, 
1.86, 2.83, 4.20, 6.14, 8.88, 12.7 (middle); $\pm$ 0.07, 0.23, 0.44, 0.71, 
1.06, 1.50, 2.08, 2.82, 3.78 (bottom).}
\label{bartours2}
\end{figure}

\vspace{1in}
\begin{figure}
\caption{The $L$ band (3.5 $\mu$m) map (top), and after modeled disk (middle)
and modeled disk and bar (bottom) subtracted. The gray contours represent 
$I_\nu<0$. Contour levels are numbered in hexadecimal, starting at 0. The 
levels (in MJy sr$^{-1}$) are 0.06, 0.22, 0.41, 0.66, 0.97, 1.37, 1.87,
2.52, 3.33, 4.37, 5.69, 7.36, 9.49, 12.2, 15.6 (top); $\pm$ 0.09, 0.31,
0.62, 1.05, 1.64, 2.46, 3.59, 5.14, 7.30, 10.3 (middle); $\pm$ 0.06, 0.21,
0.40, 0.64, 0.94, 1.33, 1.82, 2.43, 3.21 (bottom).}
\label{bartours3}
\end{figure}

\vspace{1in}
\begin{figure}
\caption{The $M$ band (4.9 $\mu$m) map (top), and after modeled disk (middle)
and modeled disk and bar (bottom) subtracted. The gray contours represent 
$I_\nu<0$. Contour levels are numbered in hexadecimal, starting at 0. The 
levels (in MJy sr$^{-1}$) are 0.06, 0.19, 0.35, 0.55, 0.80, 1.11, 1.50,
1.97, 2.56, 3.29, 4.19, 5.30, 6.68, 8.38, 10.5 (top); $\pm$ 0.08, 0.26,
0.51, 0.84, 1.28, 1.87, 2.64, 3.66, 5.02, 6.82 (middle); $\pm$ 0.06, 0.19,
0.36, 0.84, 1.17, 1.57, 2.08, 2.71 (bottom).}
\label{bartours4}
\end{figure}

\vspace{1in}
\begin{figure}
\caption{The $L$ band (3.5 $\mu$m) map after modeled disk and bar are
subtracted. The minimal mask was used in fitting the model. The gray contours 
represent $I_\nu<0$. Contour levels are numbered in hexadecimal, starting at 
0. The contour levels are $\pm$ 0.06, 0.21, 0.40, 0.64, 0.94, 1.33, 1.82, 
2.43, 3.21 MJy sr$^{-1}$.}
\label{bartour3_0}
\end{figure}

\vspace{1in}
\begin{figure}
\caption{The $J$, $K$, and $L$ surface brightness ($\log (|I_\nu|+0.001$ MJy 
sr$^{-1})$) before and after the Model S map was subtracted. The colors of this
coded-intensity image are modulated to create a countour-like effect, to better
show structure. Colors to the left of the break in the color bar (blue to dark 
green to blue) represent negative surface brightness. White pixels are 
saturated. The range is $|l|\!<\!110^\circ,\,|b|\!<\!15^\circ$. Tick-marks are 
at intervals of $10^\circ$ ($l$) and $3^\circ$ ($b$).}
\label{strips}
\end{figure}

\vspace{1in}
\begin{figure}
\caption{Three-color image of Galaxy ($|b|\!<\!45^\circ$) after Model S 
was subtracted. The modeled $J$ and $K$ bands have been scaled to have the same 
central emissivity as the $L$ band, so that all three have equal weight. 
Top: $I_\nu=-2.5$ to 2.5 MJy sr$^{-1}$. Middle: $I_\nu=-0.5$ 
to 0.5 MJy sr$^{-1}$. Bottom: $I_\nu=-0.1$ to 0.1 MJy sr$^{-1}$. Tick-marks 
are at intervals of $10^\circ$ ($l$) and $5^\circ$ ($b$). The faint pebbled
texture of the bottom panel is due to point-source residuals.}
\label{coloresid}
\end{figure}

\vspace{1in}
\begin{figure}
\caption{Logarithmic three-color image of Galaxy ($|b|\!<\!60^\circ$) before 
and after Model S was subtracted. The $J$ and $K$ bands have been scaled to 
the central emissivitity of the $L$ band. Tick-marks are at intervals of 
$20^\circ$ ($l$) and $5^\circ$ ($b$).}
\label{cobemap}
\end{figure}

\vspace{1in}
\begin{figure}
\caption{The warping of the disk, according to the three models. The dashed 
curves enclose 90\% of the fits made. The solid curves enclose the 95\% 
confidence limit on the mean elevation. The coefficients are those of a 
cubic polynomial in $R-R_w$.}
\label{warps}
\end{figure}

\vspace{1in}
\begin{figure}
\caption{The vertical density distributions of neutral hydrogen and diffuse
dust. The solid curve is the dust density from Model S, with $R_0=8.5$
kpc. The dashed curve is the H {\sc i} density from Dickey \& Lockman 
(1990).} 
\label{zdust}
\end{figure}

\vspace{1in}
\begin{figure}
\caption{Top: The log of the face-on surface brightness from Models S, E, 
and P. Our position is marked by the solar symbol. 
Bottom: The $L$-band profiles taken along the major and minor axes 
of the bar. The asterisks denote an average of bar+disk over azimuth.}
\label{faces}
\end{figure}

\vspace{0.5in}
\begin{figure}
\caption{Top: The log of the face-on surface brightness from Models S, E, 
and P, with no hole in the disk. Our position is marked by the solar 
symbol. Bottom: The $L$-band profiles taken along the major and minor 
axes of the bar. The asterisks denote an average of bar+disk over azimuth.}
\label{facenoho}
\end{figure}

\clearpage

\pagestyle{empty}

\begin{deluxetable}{lccccc} 
\tablenum{1}
\tablecaption{The quality of fit ($R_0=8.5$ kpc), using the primary mask}
\tablehead{
 \colhead{Band} & \colhead{$J$} & \colhead{$K$} & \colhead{$L$} & 
 \colhead{$M$}  & \colhead{Mean} }
\startdata
\multicolumn{6}{c}{The Disk Region} \nl
 Weight   & 0.182 & 0.364 & 0.364 & 0.091 & \nodata \nl  \tableline
MAD$_S$ (\%)   & 5.826 & 6.438 & 7.167 & 8.407 & $6.790\pm0.022$ \nl 
MAD$_E$ (\%)   & 5.871 & 6.465 & 7.226 & 8.451 & $6.815\pm0.032$ \nl 
MAD$_P$ (\%)   & 5.938 & 6.485 & 7.164 & 8.223 & $6.791\pm0.029$ \nl
$\psi_S$ & 1.199 & 1.182 & 1.106 & 1.743 & $1.209\pm0.006$ \nl  
$\psi_E$ & 1.209 & 1.180 & 1.114 & 1.749 & $1.213\pm0.005$ \nl  
$\psi_P$ & 1.223 & 1.190 & 1.100 & 1.676 & $1.207\pm0.006$ \nl
$\chi^2_S$ & 4.97 & 3.61 & 2.849 & 6.07  & $3.805\pm0.039$ \nl 
$\chi^2_E$ & 4.74 & 3.67 & 2.887 & 6.09  & $3.808\pm0.035$ \nl 
$\chi^2_P$ & 4.67 & 3.70 & 2.881 & 5.67  & $3.768\pm0.034$ \nl
\tablevspace{.15in}
\multicolumn{6}{c}{The Bar Region} \nl
 Weight   & 0.014 & 0.282 & 0.423 & 0.282 & \nodata \nl   \tableline
MAD$_S$ (\%)   & 7.53  & 4.356 & 3.856 & 4.378 & $4.197\pm0.027$ \nl 
MAD$_E$ (\%)   & 7.52  & 4.683 & 3.903 & 4.388 & $4.304\pm0.024$ \nl 
MAD$_P$ (\%)   & 8.70  & 4.744 & 4.247 & 5.192  & $4.716\pm0.028$ \nl
$\psi_S$ & 2.204 & 1.312 & 0.958 & 1.214 & $1.149\pm0.007$ \nl  
$\psi_E$ & 2.249 & 1.420 & 0.975 & 1.233 & $1.190\pm0.007$ \nl  
$\psi_P$ & 2.522 & 1.408 & 1.030 & 1.402 & $1.262\pm0.010$ \nl
$\chi^2_S$ & 16.9 & 4.91  & 2.54  & 3.53  & $3.69\pm0.05$ \nl  
$\chi^2_E$ & 16.1 & 5.54  & 2.61  & 3.70  & $3.93\pm0.06$ \nl  
$\chi^2_P$ & 20.8 & 21.7  & 19.1  & 22.2  & $20.8\pm1.3$  \nl
\enddata
\tablecomments{The mean absolute fractional deviation, MAD, is the quantity
that was minimized.}
\end{deluxetable}

\clearpage
\begin{deluxetable}{lccc} 
\tablenum{2}
\tablecaption{The parameter values ($R_0=8.5$ kpc), using the primary mask}
\tablehead{
\colhead{Parameter} & \colhead{Model S} & \colhead{Model E} & \colhead{Model P} }
\small
\startdata
Distance to G. Plane $Z_0$ \dotfill (pc) & $16.46\pm 0.18$  & $16.60\pm 0.15$& $15.95\pm0.14$ \nl
Bar Tilt Angle $\theta_0$ \dotfill($^\circ$) & $13.79\pm0.09$ & $9.52\pm0.12$& $13.18\pm0.13$ \nl
\tablevspace{.1in}
Disk Scale Length $h_r$ \dotfill(kpc) &$2.6045\pm 0.0033$  & $2.6030\pm0.0030$ & $2.567\pm0.0030$ \nl
Disk Scale Height $h_z$ \dotfill(kpc) &$0.3457\pm 0.0008$&$0.3519\pm0.0011$& $0.3467\pm0.0010$ \nl
Disk Radius $R_{max}$ \dotfill(kpc)   &$12.18\pm 0.06$   & $12.47\pm0.05$  &$12.43\pm0.07$ \nl  
Warp Line of Nodes $\phi_W$ \dotfill($^\circ$) & $0.44\pm0.09$ & $0.79\pm0.09$ &$-0.08\pm0.11$ \nl 
Disk $\rho_J(0)$ \dotfill($\mbox{MJy}\;\mbox{sr}^{-1}\:\mbox{kpc}^{-1}$)
                                   & $8.157\pm0.024$   & $8.133\pm0.032$ &$8.786\pm0.032$ \nl
Disk $\rho_K(0)$ \dotfill($\mbox{MJy}\;\mbox{sr}^{-1}\:\mbox{kpc}^{-1}$)
                                   & $6.648\pm0.022$   & $6.659\pm0.024$ &$6.970\pm0.023$ \nl
Disk $\rho_L(0)$ \dotfill($\mbox{MJy}\;\mbox{sr}^{-1}\:\mbox{kpc}^{-1}$)
                                   & $3.511\pm0.009$   & $3.449\pm0.012$ &$3.614\pm0.012$\nl
Disk $\rho_M(0)$ \dotfill($\mbox{MJy}\;\mbox{sr}^{-1}\:\mbox{kpc}^{-1}$)
                                   & $1.782\pm0.006$   & $1.738\pm0.005$ &$1.802\pm0.006$ \nl
Disk Hole Radius $O_R$ \dotfill(kpc)    & $2.973\pm0.022$   & $2.956\pm0.020$ & $3.323\pm0.018$ \nl
Disk Hole Power $O_N$ \dotfill          & $1.711\pm0.016$   & $1.595\pm0.017$ & $1.593\pm0.023$ \nl 
Hole Axis Ratio $\varepsilon$ \dotfill  & $0.8554\pm0.0042$ & $0.904\pm0.005$ & $0.939\pm0.007$ \nl  
\tablevspace{.1in}
Bar Pitch Angle \dotfill($^\circ$)       & $-0.023\pm0.027$  & $0.046\pm0.018$ & $-0.021\pm0.025$ \nl
Bar X Scale Length $a_x$ \dotfill(kpc)   & $1.696\pm0.007$   & $1.888\pm0.010$ &$1.810\pm0.009$ \nl
Bar Y Scale Length $a_y$ \dotfill(kpc)   & $0.6426\pm0.0020$ &$0.6561\pm0.0035$&$0.6450\pm0.0038$ \nl
Bar Z Scale Length $a_z$ \dotfill(kpc)   & $0.4425\pm0.0008$ &$0.4301\pm0.0020$ &$0.4324\pm0.0031$\nl
Bar Cutoff Radius $R_{end}$ \dotfill(kpc)& $3.128\pm0.014$   & $3.574\pm0.021$ & $2.713\pm0.015$ \nl
Bar Cutoff Scale Length $h_{end}$
\dotfill(kpc)                             &$0.461\pm0.005$ & $0.562\pm0.008$  & $0.882\pm0.006$ \nl
Bar Face-On Shape $C_{\perp}$ \dotfill    & $1.574\pm0.014$ & $1.609\pm0.020$ & $1.651\pm0.022$ \nl
Bar Edge-On Shape $C_{\parallel}$ \dotfill & $3.501\pm0.016$& $3.493\pm0.025$ & $3.016\pm0.023$\nl
Bar $\rho_J(0)$ \dotfill($\mbox{MJy}\;\mbox{sr}^{-1}\:\mbox{kpc}^{-1}$)
                                   & $10.52\pm0.07$   & $10.29\pm0.08$  & $11.67\pm0.06$ \nl
Bar $\rho_K(0)$ \dotfill($\mbox{MJy}\;\mbox{sr}^{-1}\:\mbox{kpc}^{-1}$)
                                   & $8.817\pm0.038$   & $8.700\pm0.048$ & $9.40\pm0.05$  \nl
Bar $\rho_L(0)$ \dotfill($\mbox{MJy}\;\mbox{sr}^{-1}\:\mbox{kpc}^{-1}$)
                                   & $4.538\pm0.017$   & $4.452\pm0.022$  & $4.880\pm0.024$ \nl
Bar $\rho_M(0)$ \dotfill($\mbox{MJy}\;\mbox{sr}^{-1}\:\mbox{kpc}^{-1}$)
                              & $2.255\pm0.009$    & $2.203\pm0.010$  & $2.388\pm0.015$ \nl 
Bar Power $n$ \dotfill        &     \nodata        & $1.4439\pm0.0049$ & $5.044\pm0.031$ \nl
Bar Core Radius $R_c$\dotfill(kpc) &   \nodata         &    \nodata    & $1.231\pm0.007$  \nl 
\tablevspace{.2in}
Dust Scale Length $h_r^d$ \dotfill(kpc)  &$3.066\pm0.019$ & $3.348\pm0.021$ & $3.425\pm0.023$ \nl
Dust Scale Height $h_z^d$ \dotfill(kpc)  & $0.1520\pm0.0008$    &
$0.1474\pm0.0009$  & $0.1465\pm0.0007$  \nl
Dust Warp Factor $x^d$ \dotfill         & $1.782\pm0.009$ & $1.755\pm0.011$ & $1.754\pm0.009$ \nl
Local Extinction $A_J$ \dotfill($J\ \mbox{mag kpc}^{-1}$)
                                   & $0.1144\pm0.0020$ & $0.1451\pm0.0027$& $0.1528\pm0.0028$ \nl
Extinction Index $\alpha$ \dotfill 
                         &$1.787\pm0.009$ & $1.762\pm0.010$ & $1.798\pm0.010$\nl
Dust Hole Radius $O_R^d$ \dotfill(kpc)  & $2.615\pm0.019$  & $2.253\pm0.017$ & $2.025\pm0.025$ \nl
Dust Hole Power $O_N^d$ \dotfill        & $2.150\pm0.022$  & $2.107\pm0.023$ & $2.463\pm0.020$  \nl 
Dust $\rho_J(0)$ \dotfill($\mbox{MJy}\;\mbox{sr}^{-1}\:\mbox{kpc}^{-1}$)
                                   & $4.642\pm0.023$   & $4.659\pm0.025$  & $4.819\pm0.029$  \nl
Dust $\rho_K(0)$ \dotfill($\mbox{MJy}\;\mbox{sr}^{-1}\:\mbox{kpc}^{-1}$)
                                   & $1.152\pm0.007$   & $1.225\pm0.008$  & $1.158\pm0.010$ \nl
Dust $\rho_L(0)$ \dotfill($\mbox{MJy}\;\mbox{sr}^{-1}\:\mbox{kpc}^{-1}$)
                                   & $2.180\pm0.014$   & $2.294\pm0.018$  & $2.437\pm0.017$  \nl
Dust $\rho_M(0)$ \dotfill($\mbox{MJy}\;\mbox{sr}^{-1}\:\mbox{kpc}^{-1}$)
                                   & $3.236\pm0.021$  & $3.142\pm0.021$  & $3.559\pm0.023$  \nl 
\enddata
\end{deluxetable}

\clearpage

\begin{deluxetable}{lccc} 
\tablenum{3}
\tablecaption{The parameter values ($R_0=8.5$ kpc), using the minimal mask}
\tablehead{
\colhead{Parameter} & \colhead{Model S} & \colhead{Model E} & \colhead{Model P} }
\small
\startdata
Distance to G. Plane $Z_0$ \dotfill (pc)  & $16.12\pm 0.19$  & $16.50\pm 0.19$& $15.66\pm0.14$ \nl
Bar Tilt Angle $\theta_0$ \dotfill($^\circ$) & $13.98\pm0.14$ & $9.84\pm0.12$ &$13.51\pm0.13$ \nl
\tablevspace{.1in}
Disk Scale Length $h_r$ \dotfill(kpc)  & $2.6009\pm 0.0040$  & $2.601\pm0.007$ & $2.567\pm0.006$ \nl
Disk Scale Height $h_z$ \dotfill(kpc)  & $0.3420\pm 0.0008$&$0.3466\pm0.0011$& $0.3440\pm0.0012$ \nl
Disk Radius $R_{max}$ \dotfill(kpc)    & $12.35\pm 0.06$   & $12.45\pm0.07$ &$12.52\pm0.08$ \nl  
Warp Line of Nodes $\phi_W$ \dotfill($^\circ$) & $0.40\pm0.11$ & $0.83\pm0.10$ &$-0.07\pm0.12$ \nl 
Disk $\rho_J(0)$ \dotfill($\mbox{MJy}\;\mbox{sr}^{-1}\:\mbox{kpc}^{-1}$)
                                   & $8.115\pm0.036$   & $8.141\pm0.043$ &$8.725\pm0.046$ \nl
Disk $\rho_K(0)$ \dotfill($\mbox{MJy}\;\mbox{sr}^{-1}\:\mbox{kpc}^{-1}$)
                                   & $6.707\pm0.029$   & $6.740\pm0.026$ &$7.000\pm0.036$ \nl
Disk $\rho_L(0)$ \dotfill($\mbox{MJy}\;\mbox{sr}^{-1}\:\mbox{kpc}^{-1}$)
                                   & $3.539\pm0.013$   & $3.478\pm0.013$ &$3.637\pm0.014$\nl
Disk $\rho_M(0)$ \dotfill($\mbox{MJy}\;\mbox{sr}^{-1}\:\mbox{kpc}^{-1}$)
                                   & $1.759\pm0.008$   & $1.724\pm0.008$ &$1.796\pm0.010$ \nl
Disk Hole Radius $O_R$ \dotfill(kpc)     & $2.912\pm0.029$   & $2.910\pm0.020$ & $3.294\pm0.025$ \nl
Disk Hole Power $O_N$ \dotfill           & $1.705\pm0.020$   & $1.572\pm0.015$ & $1.585\pm0.017$ \nl 
Hole Axis Ratio $\varepsilon$ \dotfill   & $0.822\pm0.022$   & $0.905\pm0.021$ & $0.910\pm0.014$ \nl  
\tablevspace{.1in}
Bar Pitch Angle \dotfill($^\circ$)       & $-0.05\pm0.08$   & $0.07\pm0.09$   & $0.02\pm0.08$   \nl
Bar X Scale Length $a_x$ \dotfill(kpc)   & $1.686\pm0.011$   & $1.878\pm0.011$ & $1.806\pm0.010$ \nl
Bar Y Scale Length $a_y$ \dotfill(kpc)   & $0.6429\pm0.0032$ & $0.6512\pm0.0036$ &$0.6418\pm0.0028$ \nl
Bar Z Scale Length $a_z$ \dotfill(kpc)   & $0.4420\pm0.0013$ & $0.4302\pm0.0018$ &$0.4313\pm0.0022$ \nl
Bar Cutoff Radius $R_{end}$ \dotfill(kpc)& $3.139\pm0.021$   & $3.542\pm0.019$ & $2.725\pm0.013$ \nl
Bar Cutoff Scale Length $h_{end}$
\dotfill(kpc)                             &$0.469\pm0.020$ & $0.545\pm0.019$  & $0.875\pm0.013$  \nl
Bar Face-On Shape $C_{\perp}$ \dotfill    & $1.588\pm0.017$   & $1.597\pm0.022$ & $1.655\pm0.019$ \nl
Bar Edge-On Shape $C_{\parallel}$ \dotfill &$3.466\pm0.028$  & $3.418\pm0.022$ & $2.976\pm0.023$ \nl
Bar $\rho_J(0)$ \dotfill($\mbox{MJy}\;\mbox{sr}^{-1}\:\mbox{kpc}^{-1}$)
                                   & $10.42\pm0.14$   & $10.36\pm0.09$  & $11.77\pm0.08$ \nl
Bar $\rho_K(0)$ \dotfill($\mbox{MJy}\;\mbox{sr}^{-1}\:\mbox{kpc}^{-1}$)
                                   & $8.769\pm0.045$   & $8.707\pm0.044$  & $9.40\pm0.06$  \nl
Bar $\rho_L(0)$ \dotfill($\mbox{MJy}\;\mbox{sr}^{-1}\:\mbox{kpc}^{-1}$)
                                   & $4.545\pm0.024$   & $4.465\pm0.020$  &$4.878\pm0.021$ \nl
Bar $\rho_M(0)$ \dotfill($\mbox{MJy}\;\mbox{sr}^{-1}\:\mbox{kpc}^{-1}$)
                              & $2.241\pm0.015$    & $2.180\pm0.015$  & $2.387\pm0.013$ \nl 
Bar Power $n$ \dotfill        &     \nodata        & $1.438\pm0.005$  & $5.004\pm0.031$ \nl
Bar Core Radius $R_c$\dotfill(kpc) &   \nodata         &    \nodata   & $1.224\pm0.005$ \nl 
\tablevspace{.2in}
Dust Scale Length $h_r^d$ \dotfill(kpc)  &$3.020\pm0.029$ & $3.320\pm0.026$ & $3.376\pm0.021$ \nl
Dust Scale Height $h_z^d$ \dotfill(pc)  & $0.205\pm0.006$    &
$0.2019\pm0.0048$ & $0.182\pm0.006$  \nl
Dust Warp Factor $x^d$ \dotfill         & $1.811\pm0.019$ & $1.765\pm0.014$ & $1.749\pm0.012$ \nl
Local Extinction $A_J$ \dotfill($J\ \mbox{mag kpc}^{-1}$)
                                   & $0.0898\pm0.0033$ & $0.1116\pm0.0038$& $0.1236\pm0.0039$ \nl
Extinction Index $\alpha$ \dotfill 
                         &$1.987\pm0.023$ & $2.011\pm0.024$ & $1.979\pm0.020$\nl
Dust Hole Radius $O_R^d$ \dotfill(kpc)  & $2.684\pm0.032$  & $2.222\pm0.020$ & $2.051\pm0.026$ \nl
Dust Hole Power $O_N^d$ \dotfill        & $2.182\pm0.023$  & $2.116\pm0.012$ & $2.466\pm0.020$ \nl 
Dust $\rho_J(0)$ \dotfill($\mbox{MJy}\;\mbox{sr}^{-1}\:\mbox{kpc}^{-1}$)
                                   & $4.681\pm0.028$   & $4.699\pm0.034$  & $4.795\pm0.036$  \nl
Dust $\rho_K(0)$ \dotfill($\mbox{MJy}\;\mbox{sr}^{-1}\:\mbox{kpc}^{-1}$)
                                   & $1.146\pm0.008$   & $1.229\pm0.008$  & $1.145\pm0.010$ \nl
Dust $\rho_L(0)$ \dotfill($\mbox{MJy}\;\mbox{sr}^{-1}\:\mbox{kpc}^{-1}$)
                                   & $2.196\pm0.018$   & $2.320\pm0.024$  & $2.455\pm0.019$  \nl
Dust $\rho_M(0)$ \dotfill($\mbox{MJy}\;\mbox{sr}^{-1}\:\mbox{kpc}^{-1}$)
                                   & $3.185\pm0.031$  & $3.139\pm0.025$  & $3.559\pm0.026$  \nl 
\enddata
\end{deluxetable}

\clearpage
\begin{deluxetable}{lccccc} 
\tablenum{4}
\tablecaption{Model S vs $R_0$, using the primary mask}
\tablehead{ \colhead{Distance to G. Center $R_0$ (kpc)} &\colhead{7.5} &
\colhead{8.0} & \colhead{8.5} & \colhead{9.0} & \colhead{9.5} }  
\small
\startdata
Distance to G. Plane $Z_0$ \dotfill (pc)&14.90 & 16.08 & 16.46 & 16.27 &16.69\nl
Bar Tilt Angle $\theta_0$ \dotfill($^\circ$)&13.69& 12.86 &13.79&14.58&15.02\nl
\tablevspace{.1in}
Disk Scale Length $h_r$ \dotfill(kpc)&2.3504 & 2.476 & 2.6045 &2.7092&2.8305\nl
Disk Scale Height $h_z$ \dotfill(kpc)& 0.3196 &0.3340 &0.3457& 0.3602&0.3719 \nl
Disk Radius $R_{max}$ \dotfill(kpc)  & 11.61 & 12.11 & 12.18  &12.87 &13.35\nl  
Warp Line of Nodes $\phi_W$ \dotfill($^\circ$)& 0.51 & 0.48 &0.44 &0.52&0.48\nl 
Disk $\rho_J(0)$ \dotfill($\mbox{MJy}\;\mbox{sr}^{-1}\:\mbox{kpc}^{-1}$)
                                   & 8.213 & 8.272 & 8.157 & 8.144 &8.088 \nl
Disk $\rho_K(0)$ \dotfill($\mbox{MJy}\;\mbox{sr}^{-1}\:\mbox{kpc}^{-1}$)
                                   & 6.725 & 6.731 & 6.648 & 6.666&6.671\nl
Disk $\rho_L(0)$ \dotfill($\mbox{MJy}\;\mbox{sr}^{-1}\:\mbox{kpc}^{-1}$)
                                   & 3.586 & 3.527 & 3.511 & 3.513 &3.506\nl
Disk $\rho_M(0)$ \dotfill($\mbox{MJy}\;\mbox{sr}^{-1}\:\mbox{kpc}^{-1}$)
                                   & 1.806 & 1.781 & 1.782 & 1.745 &1.746 \nl
Disk Hole Radius $O_R$ \dotfill(kpc) & 2.733 & 2.939 & 2.973 & 2.971&2.998 \nl
Disk Hole Power $O_N$ \dotfill       & 1.729 & 1.721 & 1.711 & 1.784&1.796 \nl 
Hole Axis Ratio $\varepsilon$ \dotfill &0.855 &0.870 &0.8554 & 0.955&0.849 \nl  
\tablevspace{.1in}
Bar Pitch Angle \dotfill($^\circ$)    & -0.01 & 0.036 &-0.023 & -0.159 &-.170\nl
Bar X Scale Length $a_x$ \dotfill(kpc)& 1.581 & 1.693 & 1.696  & 1.763 &1.817\nl
Bar Y Scale Length $a_y$ \dotfill(kpc)&0.5736 &0.6266 &0.6426 &0.6734 &0.7102\nl
Bar Z Scale Length $a_z$ \dotfill(kpc)&0.3970 &0.4172 &0.4425 &0.471 &0.4966\nl
Bar Cutoff Radius $R_{end}$ \dotfill(kpc)&2.631 &2.967 &3.128  &3.315 &3.509\nl
Bar Cutoff Scale Length $h_{end}$
\dotfill(kpc)                          &0.451 &0.440 &0.461 &0.445 &0.509\nl
Bar Face-On Shape $C_{\perp}$ \dotfill &1.634 &1.550 &1.574 &1.516 &1.478\nl
Bar Edge-On Shape $C_{\parallel}$ \dotfill &3.446 &3.427 &3.501 &3.352 &3.279\nl
Bar $\rho_J(0)$ \dotfill($\mbox{MJy}\;\mbox{sr}^{-1}\:\mbox{kpc}^{-1}$)
                                   & 10.75 & 10.65 & 10.52 & 10.48 &10.38\nl
Bar $\rho_K(0)$ \dotfill($\mbox{MJy}\;\mbox{sr}^{-1}\:\mbox{kpc}^{-1}$)
                                   & 8.942 & 8.849 & 8.817 & 8.79 &8.738 \nl
Bar $\rho_L(0)$ \dotfill($\mbox{MJy}\;\mbox{sr}^{-1}\:\mbox{kpc}^{-1}$)
                                   & 4.676 & 4.581 & 4.538 & 4.484 &4.470\nl
Bar $\rho_M(0)$ \dotfill($\mbox{MJy}\;\mbox{sr}^{-1}\:\mbox{kpc}^{-1}$)
                              & 2.360 & 2.286 & 2.255 & 2.234 &2.199 \nl 
\tablevspace{.2in}
Dust Scale Length $h_r^d$ \dotfill(kpc)&2.688 &2.981 &3.066 &3.249 &3.387\nl
Dust Scale Height $h_z^d$ \dotfill(kpc)&0.1575 &0.1500 &0.1520 &0.180 &0.194\nl
Dust Warp Factor $x^d$ \dotfill        &1.787 &1.796 &1.782 &1.778 &1.757 \nl
Local Extinction $A_J$ \dotfill($J\ \mbox{mag kpc}^{-1}$)
                                   &0.0656 &0.1050 &0.1144 &0.1209 &0.1311 \nl
Extinction Index $\alpha$ \dotfill 
                         & 1.887 & 1.783 & 1.787 & 1.740 & 1.696 \nl
Dust Hole Radius $O_R^d$ \dotfill(kpc)  &2.624 &2.585 &2.615 &2.623 &2.636\nl
Dust Hole Power $O_N^d$ \dotfill        &2.131 &2.102 &2.150 &2.093 &2.102\nl 
Dust $\rho_J(0)$ \dotfill($\mbox{MJy}\;\mbox{sr}^{-1}\:\mbox{kpc}^{-1}$)
                                 &4.702 &4.652 &4.642 &4.614 &4.625\nl
Dust $\rho_K(0)$ \dotfill($\mbox{MJy}\;\mbox{sr}^{-1}\:\mbox{kpc}^{-1}$)
                                   & 1.147 & 1.161 & 1.152 & 1.135 &1.131 \nl
Dust $\rho_L(0)$ \dotfill($\mbox{MJy}\;\mbox{sr}^{-1}\:\mbox{kpc}^{-1}$)
                                   & 2.244 & 2.187 & 2.180 & 2.157 &2.148 \nl
Dust $\rho_M(0)$ \dotfill($\mbox{MJy}\;\mbox{sr}^{-1}\:\mbox{kpc}^{-1}$)
                                   & 2.360 & 3.213 & 3.236 & 3.179 &3.148 \nl 
\tablevspace{.2in}
Total Luminosity \dotfill ($\times10^{10}\;L_\odot$)&1.9&2.1&2.3&2.7&3.0\nl
\enddata
\tablecomments{The emissivities $\rho$ are given in uncommon units, but can be 
converted: $1\;\mbox{MJy}\;\mbox{sr}^{-1}\:\mbox{kpc}^{-1}=9.522\times10^9\;
\mbox{W}\;\mbox{pc}^{-3}\:\mbox{Hz}^{-1}\:\mbox{sr}^{-1}$.}
\end{deluxetable}

\end{document}